\documentstyle[12pt]{article}
\begin{document}

\vspace{3mm}
\noindent
{\Large \bf Another Exact Inflationary Solution}\\\\
{\large Adam T. Kruger and John W. Norbury}\\
{\em (Physics Department, University of Wisconsin-Milwaukee,
P.O. Box 413, Milwaukee, Wisconsin 53201, USA.)}

\vspace{8mm}
\noindent
{\bf Abstract}

\noindent
A new closed-form inflationary solution is given for a hyperbolic
interaction potential.  The method used to arrive at this solution is
outlined as  it appears possible to generate additional
sets of equations which satisfy the model. In addition a new form of decaying cosmological constant is
presented.
\vspace{3mm}

\vspace{4mm}

\section {Introduction}

 Exact inflationary solutions 
\cite{ Barrow87, Lucchin, Barrow94, Schunck, Islam, Zhuravlev} are of interest because they allow one to
study physical effects in a simple manner than would otherwise be allowed from numerical studies. In
addition, they allow one to write down exact forms for decaying cosmological constant models
\cite{Norbury-plb} as we shall see below. It is thus always interesting if one can think of new methods
for developing exact solutions to inflation models.
The results of the present work are threefold. Firstly a new method for generating exact inflationary
solutions is given. Secondly a new inflation solution is presented.  Thirdly, from this exact
solution, a new form of decaying cosmological constant is presented.

Exact inflationary solutions have
been studied before 
\cite{ Barrow87, Lucchin, Barrow94,  Schunck, Islam,
Zhuravlev, Starkovich, Lidsey, Ellis, Salopek, Muslimov} and we would like to
point out the difference between previous work and the present work. It is well known that inflationary
potentials of the form $V(\phi) \propto e^{\pm \phi}$ can be solved exactly and these types of models
were studied in \cite{Islam, Starkovich, Salopek, Muslimov}.
 This type of potential is considered in the present work  (Section 3.1), but only 
as a check on the more
 general solution method that we develop. The works by Lidsey \cite{Lidsey}, Ellis and
Madsen \cite{Ellis} and Lucchin and Matarrese \cite{Lucchin} consider more complicated potentials and
 are all similar in that they specify
either  $H \equiv \frac{\dot a}{a}$ \cite{Lidsey}
or $a(t)$
\cite{Lucchin,Ellis} initially and {\em then} go on to derive the corresponding $V(\phi)$. 
Ideally however, one would wish to do this the other way around, i.e. specify a form for  $V(\phi)$
initially and then find the corresponding expressions for $a(t)$ and $\phi (t)$. Our present work does
not quite achieve this, but we come closer to this goal than   previous work \cite{ Barrow87, Lucchin,
Barrow94, Schunck, Islam, Zhuravlev, Starkovich, Lidsey, Ellis, Salopek, Muslimov}.
We specify a function $F(\phi)$ (see later) which gives  $V(\phi)$. {\em Then} we are able to solve for 
$a(t)$ and $\phi (t)$. As mentioned above the other works \cite{Lucchin, Lidsey, Ellis} do this the
other way around, which is less satisfactory.

The inflationary scenario holds that at a very early time
the universe underwent a rapid period of quasi-exponential expansion.  This idea 
was proposed as a way of eliminating the necessity of fine-tuning of
initial conditions in the standard hot big bang model, as this rapid
expansion would account for the flatness of the observed universe and also
explain why regions of the universe that appear causally disconnected are
in thermal equilibrium \cite{Kolb}.  The model that is studied here
assumes the interaction of a classical scalar (inflaton) field.  It also
assumes that on large scales the universe has a uniform energy density and
that the structure is the same in every direction. These assumptions of
homogeneity and isotropy and the equations of General Relativity give rise
to the so called Friedmann models which seem to describe the large-scale
behavior of our universe.  These models allow for the inclusion of a
cosmological constant term $\Lambda$, however in this study it
is assumed that this term is identically zero and that the expansion is
driven by interaction with the scalar field (i.e. that the energy density
of the scalar field can act as an effective cosmological term).

The size of the universe is given in terms of the scale factor $a$.   In the
Friedmann-Robertson-Walker cosmological model the evolution of the scale
factor $a$ is specified by \cite{MTW, Norbury-ejp}
\begin{equation}
H^{2}\equiv \left(\frac{\dot{a}}{a}\right)^{2}=\frac{8\pi G}{3}\rho-\frac{k}{a^{2}}+
\frac{\Lambda}{3},
\label{1-1}
\end{equation}
where $H$ is the Hubble parameter and overdots indicate time derivatives and $\rho$ is the energy
density. The curvature term $\frac{k}{a^{2}}$ will be neglected in this model for
the sake of simplicity and also because this term will quickly approach
zero for any inflationary solution. In models whose only matter source is
a scalar field $\phi$ the energy density has the form
\begin{equation}
\rho_{\phi} = \frac{1}{2} \dot{\phi}^{2}+V,
\label{1-2}
\end{equation}
where $V \equiv V(\phi)$ is the interaction potential.
During a period where the energy density of the scalar field dominates,
equation (\ref{1-1}) takes the form
\begin{equation}
H^{2}\equiv\left(\frac{\dot{a}}{a}\right)^{2}=\frac{8 \pi G}{3}\left(\frac{1}{2}\dot{\phi}^2+V\right)
\label{1-3}
\end{equation}
giving the differential equation determining $a(t)$ as
\begin{equation}
H \equiv \frac{\dot{a}}{a}= \pm \frac{g}{3}\sqrt{\dot{\phi}^2+2 V}
\label{1-3a}
\end{equation}
 where  
\begin{equation}
g \equiv \sqrt{12\pi G}
\label{1-3b}
\end{equation}
The $\pm$ signs in (\ref {1-3a}) indicate an expanding or contracting
solution for $a(t)$.  The equation which specifies the time evolution of the field
density is
\begin{equation}
\ddot{\phi}+3 H \dot{\phi}+ V^\prime =0
\label{1-4}
\end{equation}
with $V^\prime \equiv \frac{dV}{d\phi}$. Equations (\ref{1-3a}) and (\ref{1-4}) form a set of coupled
equations which specify the expansion in this model. Eliminating
$H \equiv \frac{\dot{a}}{a}$
 yields 
\begin{equation}
\ddot{\phi} + g \dot{\phi}\sqrt{\dot{\phi^2}+2V}+V^\prime =0
\label{1-5}
\end{equation}
where we have included only the expanding solution from (\ref{1-3a}). The contracting solution would
involve a $-$ sign in front of the square root in the above equation. In principle one could specify a
potential
$V
\equiv V(\phi)$ and solve this equation for $\phi(t)$.  One could then find the functional
form of all other dynamical quantities, i.e. $a(t)$, $\rho(a)$, etc. \cite{Norbury-plb}.  In
practice this is difficult and few exact solutions have been found.

\section {General Solution}

In this paper a new closed-form solution is presented and the method
of solution is described in detail as it may provide opportunities
to find additional exact solutions. The solution  is obtained by finding equivalent coupled equations
that are written in terms of an arbitrary function $F$.  One can specify
$F(\phi)$ to be any convenient function. This will determine the functional form
of the interaction potential, which together with $F$,
 determine the scalar field density $\phi(t)$.

Defining the function $F \equiv F(\phi)$ via
\begin{equation}
\dot\phi \equiv \pm \sqrt{( F -1)V}
\label{1-10}
\end{equation}
gives equation (\ref{1-5}) as
\begin{equation}
\ddot{\phi} \pm g V\sqrt {F^2-1}+V^\prime  =0
\label{1-11}
\end{equation}
 Equations (\ref{1-10}) and
(\ref{1-11}) form a set of coupled equations which are equivalent to equation
(\ref{1-5}).  They can be uncoupled by differentiating (\ref{1-10}) with respect
to time and substituting for
$\ddot{\phi}$ in (\ref{1-11}). Solving for $\ddot{\phi} = \frac{1}{2} \frac{d \dot \phi^2}{d \phi}$
from (\ref{1-10}) we have
\begin{equation}
\ddot{\phi}=\frac{1}{2}\left[(  F-1) V^\prime  +  V F^\prime \right]
\label{1-12}
\end{equation}
(with $ F^\prime \equiv \frac{dF}{d\phi}$ ) and inserting this expression for $\ddot{\phi}$ into
(\ref{1-11}) leads to
\begin{equation}
( F+1) V^\prime  + V F^\prime \pm 2 g V\sqrt {F^2-1} = 0
\label{1-13}
\end{equation}
It would be nice if we could just specify the interaction potential $V(\phi)$ and solve this first-order
differential equation for $F(\phi)$. However in the absence of such a solution, we instead choose the
less convenient method of doing it the other way around, i.e. by specifying $F(\phi)$ and then solving
for $V(\phi)$. It can be seen that if one chooses $F \equiv F(\phi)$,
then equation (\ref{1-13}) is always separable and the potential is given
by
\begin{equation}
V =\beta \exp  \left( - \int   \frac{  F^\prime \pm 2 g \sqrt
{F^2-1} }{ F +1}d\phi \right)
\label{1-21-1}\
\end{equation}
with $F \equiv F(\phi)$ and $\beta$ is a constant. This may be simplified to 
\begin{equation}
V = B  \exp  \left( \mp  2 g \int  \sqrt{ \frac{F-1 }{F+1}}
d\phi \right)
\label{1-21}\
\end{equation}
where $B$ is a constant. 
It is important to note that equations
(\ref{1-10}), (\ref{1-11}), (\ref{1-13}), (\ref{1-21-1}), (\ref{1-21}) are  consistent {\em only} if one
takes either {\em all} the upper signs or {\em all} the lower signs. One cannot take upper signs in one
equation and lower signs in another.

The choice of function $F$ is arbitrary. 
 When one specifies $F(\phi)$
then one has the potential $V(\phi)$ as given in the above equation and also the solution for $\phi(t)$
and
$a(t)$ as given in equations (\ref{1-10}) and (\ref{1-3a}). In other words one has a general method of
finding exact inflationary solutions.
This expression (\ref{1-21}) for $V(\phi)$ and the procedure for getting
$\phi (t)$  by solving (\ref{1-10}) is the first main result of the present paper.

\section {Special Cases}

\subsection{$F =  constant$}

 It should
be noted, for example, that if one chooses $F=constant$, then one is led to
the solution for $\phi (t) $ given by Barrow \cite{ Barrow87, Lucchin, Barrow94}, although the
constants could be chosen differently.  Equation(\ref{1-21}) gives
\begin{equation}
V=B \exp \left( \mp 2 g \sqrt{\frac{F-1}{F+1}}  \phi \right)
\label{1-19}
\end{equation}
with solutions  from (\ref{1-10})  of the form
\begin{equation}
\phi(t)= 
\pm \frac{1}{g} \sqrt{\frac{F+1}{F-1}} \ln \left[ \pm \frac{g (F-1)\sqrt{B}}{\sqrt{F+1}}(t -
C)\right]
\label{1-20}
\end{equation}
with  $C$ being a constant and the upper or lower signs in (\ref {1-20}) in front of the term
$\frac{1}{g}$ correspond to the upper or lower signs in (\ref{1-19}) or (\ref{1-10}). 
However the $\pm$ signs occurring in the square brackets occur {\em regardless} of what sign is taken
in (\ref{1-19}) or (\ref{1-10}). That is, (\ref{1-19}) or (\ref{1-10}) have two solutions for {\em
each} choice of sign, the two solutions corresponding to the two signs in the square brackets.
Equation (\ref{1-20}) is simply the
Barrow solution
\cite{Barrow87, Lucchin}. One can now easily verify that (\ref{1-20}) and (\ref{1-19}) satisfy
(\ref{1-5}). 

Substituting (\ref{1-19}) and (\ref{1-20}) into (\ref{1-3a})  yields
\begin{equation}
a(t) \propto t^{\frac{1}{3}\left(\frac{F+1}{F-1}\right)}
\label{1-200}
\end{equation}
and eliminating $\phi(t)$ in favor of $t$
using (\ref{1-20}) gives, from (\ref{1-2}),  
\begin{equation}
\rho (a) \propto \frac{1}{a^{6 \left(\frac{F-1}{F+1}\right)}}
\label{1-201}
\end{equation}
which is of the form of a decaying cosmological constant \cite{Norbury-plb}. The direct relation
between $V(\phi)$ and $\rho(a)$ is perhaps made clearer if we define a constant $\lambda$ via
\begin{equation}
V(\phi) \equiv B \exp (-\lambda \phi)
\label{1-202}
\end{equation}
which, upon comparing to  (\ref{1-19}) is
\begin{equation}
\lambda \equiv 2 g \sqrt{\frac{F-1}{F+1}}
\label{1-203}
\end{equation}
so that
\begin{equation}
a(t) \propto t^{\frac{4 g^2}{3 \lambda^2}}
\label{1-204}
\end{equation}
giving
\begin{equation}
\rho(a) \propto \frac{1}{a^{\frac{3\lambda^2}{2 g^2}}}
\label{1-205}
\end{equation}
which is consistent with equation (37) of reference \cite{Norbury-plb}.
If $\lambda$ is determined by say some other particle physics theory then $\rho(a)$ is specified. In
the next section we pick $\lambda = g$ for simplicity. Doing that here would give

\begin{equation}
a(t) \propto t^{\frac{4}{3}}
\label{1-206}
\end{equation}
and
\begin{equation}
\rho(a) \propto \frac{1}{a^{\frac{3}{2}}}
\label{1-207}
\end{equation}
thus showing that the Barrow solution results in a {\em power law} decaying cosmological constant. Such
a power law decay is the assumption in many decaying cosmological constant models.

\subsection{$F = \cosh (\lambda \phi)$}

One of the other main results of the present paper is the exact solution for $\phi (t)$ resulting from
the choice $F\equiv \cosh(\lambda \phi)$.  Then equation (\ref{1-21}) gives
\begin{equation}
V(\phi) = C \left( 1 + \cosh \lambda \phi \right)^{\mp 2 g^\prime - 1}
\label{1-20ab}
\end{equation}
where $C$ is a constant and with the $\pm$ sign corresponding to the $\pm$ sign in (\ref{1-10}) and the
expanding solution (\ref{1-5}) and
\begin{equation}
g^\prime \equiv \frac{g}{\lambda}
\label{1-20abc}
\end{equation}
Thus equation (\ref{1-10}) becomes
\begin{equation}
\dot \phi = \pm \sqrt{C} \frac{\sinh \lambda \phi}{(1 + \cosh
\lambda \phi)^{1 \pm  g^\prime}}
\label{1-20abc}
\end{equation}
This equation is only consistent if all upper or all lower signs are taken, i.e. one should not mix
upper and lower signs. This equation  is most easily solved if the denominator disappears,
i.e. the constant $\lambda$ is chosen so that $\lambda = g \equiv
\sqrt{12 \pi G}  $
 and with the {\em lower signs}, corresponding to the
lower sign in definition (\ref{1-10}).  {\em Thus for the rest of this paper, whenever $\lambda$ appears
it is taken to be
$\lambda = g$. }
 The upper $-$ sign in equation(\ref{1-20ab}), corresponding to the upper $+$ sign in definition
(\ref{1-10}) gives
\begin{equation}
V(\phi)=\frac{C}{(1+ \cosh \lambda \phi)^3}
\label{1-111}
\end{equation}
and the {\em lower} $+$ sign in (\ref{1-20ab})  corresponding to the lower $-$ sign in definition
(\ref{1-10}) gives
\begin{equation}
V(\phi)= C (1+ \cosh \lambda \phi). 
\label{1-111a}
\end{equation}
We will study this latter solution (\ref{1-111a}) whose
 graph  of $V(\phi)$ versus $\phi$ looks similar to a parabola.
Using this result (\ref{1-111a})
in equation (\ref{1-10}) leads to
\begin{equation}
\dot{\phi}= - \sqrt {C} \sinh \lambda \phi
\label{1-16}
\end{equation}
Solving for $\phi(t)$ yields
\begin{eqnarray}
\phi(t)=\frac{2}{\lambda} \coth^{-1} \left\{\exp \left[ \lambda  \sqrt {C} (t -
D)\right]\right\}
\nonumber
\\ =\frac{1}{\lambda}\ln \left\{ \frac{\exp \left[  \lambda \sqrt {C} (t -
D)\right] + 1}{\exp
\left[ \lambda 
\sqrt { C} (t - D)\right] -1} \right\}
\label{1-17}
\end{eqnarray}
where $D$ is a constant. The form of the potential suggests that
$\phi(t)$ be a function that decreases from an initial maximum value similar to the chaotic inflation
model.  We can choose $D=0$ and so (\ref{1-17})  can be written as
\begin{equation}
\phi(t)= \frac{1}{\lambda}\ln \left[\frac{\exp(\lambda\sqrt{C} t) +
1}{\exp( \lambda \sqrt{C} t) - 1}\right].
\label{1-18}
\end{equation}
This expression for $\phi(t)$ is just a monotonically
decreasing function of $t$. Again one can easily verify that (\ref{1-18}) and (\ref{1-111a})
satisfy (\ref{1-5}).

The form of the potential (\ref{1-111a})
is similar to the chaotic inflation potential,
$V(\phi)=\frac{1}{2}m^2\phi^2$,
in that the initial value of the scalar field determines how the expansion
proceeds and also that the true vacuum state (global potential minimum) is
located at $\phi = 0$ \cite{Linde}. For the chaotic inflation potential
the value of the scalar field oscillates around this minimum with
$\phi(t)$ taking negative values.  The hyperbolic potential given here
does not lead to this behavior, instead the scalar field (\ref{1-18}) smoothly
approaches zero at late times.

This solution can be used to determine the evolution of the scale factor
with time by solving equation (\ref{1-3a}) for the expanding ($+$ square root) solution.
Substituting (\ref{1-111a}) and (\ref{1-16}) into (\ref{1-3a})  yields
\begin{equation}
a(t) = \left[\exp(2 \lambda\sqrt {C}\hspace{1.2mm} t)-1\right]^{\frac{1}{3}},
\label{1-22}
\end{equation}
for $a(0)=0$. 

Equations (\ref{1-111a}), (\ref{1-18}) and (\ref{1-22}) constitute the second main result of the present
paper, in that a new exact inflationary solution has been given for the potential specified by  
$V(\phi)=C(1+\cosh\lambda\phi)$.

Many authors
have studied models where the cosmological term decays as the scale factor
increases and have focused on examples where the functional relationship
is one of power-law decay with one or more terms \cite{Norbury-plb,Ozer}.
Here the effective cosmological term arises from the energy density of the
scalar field. From equations (\ref{1-2}) and (\ref{1-111a},\ref{1-18},\ref{1-22}) we
obtain a functional form for $\rho(a)$  namely
\begin{equation}
\rho(a)=2C \left(1 +\frac{2}{a^3}+\frac{1}{a^6}\right)
\label{1-22a}
\end{equation}
(Note that $\lambda$ does not appear in this expression.) 
 This is the third main result of the present paper presenting a new form of decaying cosmological
constant. This differs from previous models \cite{Ozer} in that a {\em constant} is added onto
{\em two} power law decay terms.

\newpage
\section{Summary}

In summary, we have introduced a general formula (\ref{1-21}) for finding exact inflationary
solutions. This formula has been checked by showing that it reproduces Barrow's solution
\cite{Barrow87}. Secondly, a  new exact solution for a hyperbolic interaction potential has been
presented. Other exact solutions can be found by choosing different functional forms for $F$ appearing in
(\ref{1-21}), and as such should prove useful in studying other inflationary models. Finally a new
decaying cosmological constant model, which includes a constant term plus two power law decay terms,
results from this exact solution.\\

\noindent
{\bf Acknowledgements}\\
ATK is funded by the Wisconsin Space Grant Consortium and NSF grant PHY-9507740.

\end{document}